# The Fabra-ROA Telescope at Montsec (TFRM): A Fully Robotic Wide-field Telescope for Space Surveillance and Tracking

## European Space Surveillance Conference
## 7-9 June 2011

Montojo F.J.[1], Fors O.[2,3], Muiños J.L.[1], Núñez J.[2,3], López-Morcillo R.[1], Baena R.[2,3], Boloix J.[1], López-Moratalla T.[1], Merino M.[2,3]

[1]Real Instituto y Observatorio de la Armada (ROA)
Plaza de las Marinas s/n, San Fernando
E-11110 Cádiz (SPAIN)
fjmontojo@roa.es

[2]Observatori Fabra, Reial Acadèmia de Ciències I Arts de Barcelona
Camí de l'Observatori s/n , Barcelona
E-08002 Barcelona (SPAIN)

[3]Dep. D'Astronomia i Meteorologia i Institut de Ciències del Cosmos (ICC), Universitat de Barcelona (UB/IEEC)
Martí i Franquès 1, Barcelona
E-08028 Barcelona (SPAIN)

**ABSTRACT**

Since the beginning of the Space Age optical sensors have been one of the main instruments for positioning and tracking known space objects. Nowadays, the unrelenting growth of man-made objects together with the overcrowding of the useful satellite orbits, and the real space debris and NEO hazards, has made necessary to carry out surveys of the space looking for uncatalogued objects. Optical telescopes play a key role in the Space Surveillance and Tracking (SST) as a primary Space Situational Awareness element and, it is known, that the best instrument for this task is a fully robotic wide-field telescope with a minimum aperture of 40cm.

The Baker-Nunn Cameras (BNCs) were produced by the Smithsonian Institution during the late 50s as an optical tracking system for artificial satellites. These wide-field telescopes of 50cm of aperture were manufactured by Perkin-Elmer (optics) and Boller & Chivens (mechanics) with the highest quality specifications. The TFRM is a fully robotic refurbished BNC that exploits the excellent mechanical and optical original design to obtain an equatorial telescope with a useful 4.4º x 4.4º CCD field of view. TFRM is therefore a European asset very well suited for SST (satellites and space debris) and NEO observations. Moreover, its control system allows tracking of all kinds of orbits, including LEOs.

**THE ORIGIN OF BAKER-NUNN CAMERAS**

The Baker-Nunn Cameras (BNCs) were designed and constructed by the Smithsonian Institution in the early Space Age as the North American solution for tracking the first artificial satellites. The USA placed 21 of these BNCs all over the world to have global coverage and in 1958 one of them was installed at the Spanish Navy Observatory (ROA), in San Fernando (Cádiz), southern Spain.

The BNCs were optically manufactured by Perkin-Elmer as an f/1 0.5m photographic wide field telescope, 30º x 5º field of view (FoV), with a spot size smaller than 20μm throughout the field and mechanically manufactured by Boller & Chivens, with extremely high optical and mechanical specifications.

During the 80s, the BNCs were superseded by newer technologies (laser, radar and CCDs) and the San Fernando BNC (Fig.1) was donated to ROA, where it has been maintained inactive but in excellent state of conservation.

Today, space debris has become a real threat that demands to come back again to wide-field optical sensors capable to perform sky surveys not only to follow up catalogued (known) objects, as the BNCs were used for, but also to look for unknown objects.

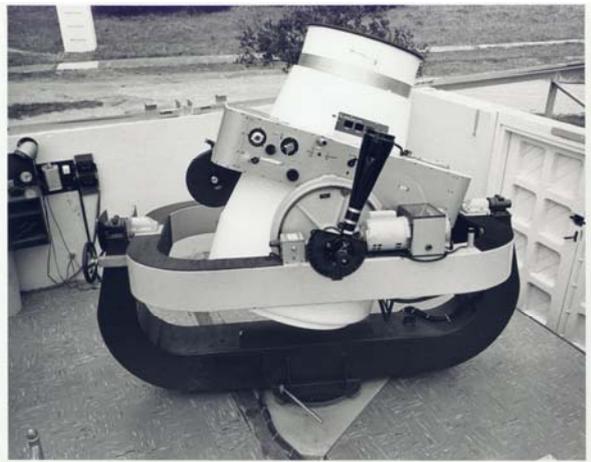 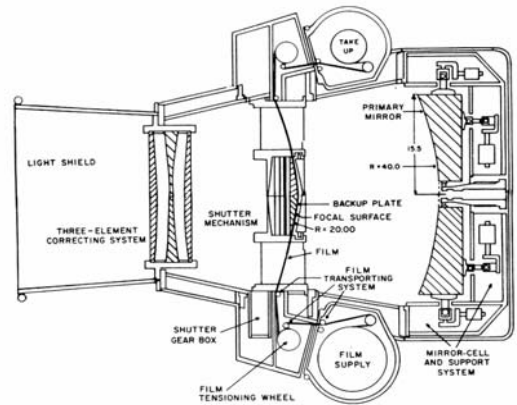

Fig.1. Original Baker-Nunn Camera installed at San Fernando Observatory site. On the right a technical drawing. Note the three-element lenses corrector cell, the curved photographic film and the primary mirror with its counterweights system.

**REFURBISHMENT PROJECT**

The BNC was refurbished by a project carried out by the Reial Acadèmia de Ciències i Arts de Barcelona (RACAB) jointly with the Real Instituto y Observatorio de la Armada (ROA). Both institutions were founded in the mid eighteenth century and still continue their activities in different scientific fields. Specifically, with the commissioning of this refurbished telescope, the two institutions that jointly manage this facility, have begun a new phase in the Astronomy field (http://www.am.ub.es/bnc).

The refurbishment process can be summarized in the following steps:

- The original alt-azimuthal mount has been converted into equatorial type. The motion of Hour Angle (HA) and Declination (DEC) axes are geared with digital servo drives and feeded-back with absolute position encoders.
- A new spider vanes and focus system for the CCD was designed and machined by Moreno Pujal S.L. (Fig. 2). Such support is athermal and the focus accuracy was found to be ±10μm.
- A 20μm spot size has been guaranteed throughout a flat field FoV of 6.25º in diameter thanks to the design (Malcolm J. MacFarlane, PhD.) and the manufacture of a CaF 64mm field flattener (Harold Johnson Optical Labs, Inc.) and a 180mm fused silica ellipsoidal meniscus (Tucson Optical Research Corporation, Inc.) lenses.
- By the first-time a large-format chip commercial CCD with a field flattener lens inside the camera body was designed and manufactured by Finger Lakes Instrumentation, Inc. (Fig.3). The flattener lens was placed as close as 0.65mm to the CCD chip. The camera also includes a Schott GC475 colour filter as a camera window, a new large aperture (90mm) shutter and a glycol recirculation cooling system.
- A 12mx5mx4.5m reinforced glass-fiber enclosure was designed and constructed by GRPro, Inc (Fig.4). This prototype is inspired in the one built for Super-WASP project [1]. It has turned to be very robust to weather conditions with no incidence to their mobile parts (sliding roof and South downloadable end wall).
- An state-of-the-art observatory control software based on the Instrument-Neutral Distributed Interface (INDI) device communication protocol, was designed and implemented by Elwood C. Downey [2]. It allows both remote and robotic control of every device of the observatory via Internet with the use of Java clients or schedulers (Fig.5). Observing blocks are directly written in XML. Complex environmental conditions decisions can be easily performed at user-level by means of high-level scripting languages such as Perl, Python, Bash, etc.
- Mirror realuminization and repolishing of outermost 50cm lens were performed by H.L.Clausing, Inc. and Harold Johnson Optical Labs, Inc., respectively.

Once the refurbishment process was completed, the main features of the telescope are: equatorial mount, a three 0.5m aperture lenses corrector cell and a 0.8m diameter mirror with an f/0.96 focal ratio. A FLI Proline 16803 glycol cooled camera with a 4k x 4k format CCD of 9μm pixel size, resulting in a 4.4º x 4.4º useful FoV with a pixel scale of 3.9 arcseconds. The motorized Hour Angle and Declination axes are moved by digital servo drives at arbitrary rates with the feedback of absolute encoders. It is remarkable that the entire new system is based on commercially available components except the telescope itself, of course.

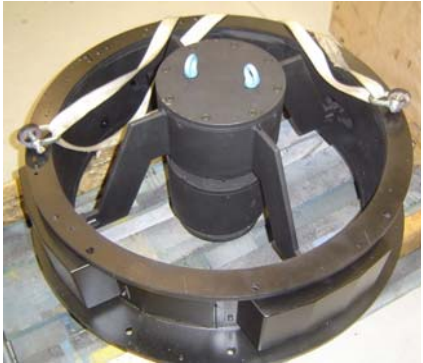

Fig.2. Spider vanes and focus support.

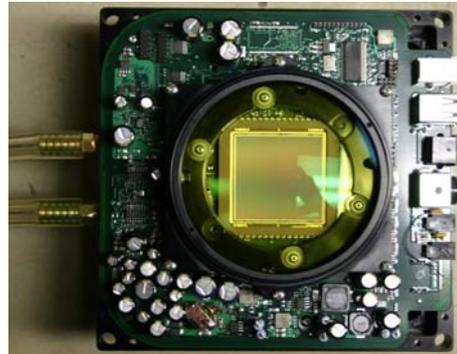

Fig.3. Finger Lakes Instrumentation, Inc. CCD with flattener lens and glycol cooling.

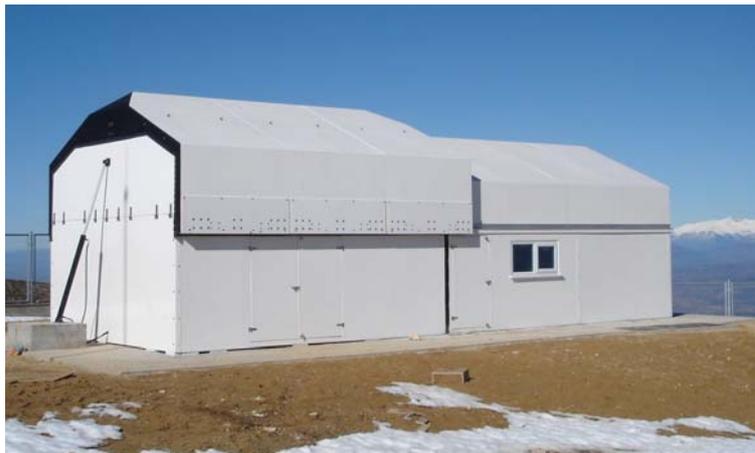

Fig.4. Reinforced glass-fiber enclosure designed and constructed by GRPro, Inc. Note the North-South sliding roof and the South downloadable end wall which allows observing at low altitudes.

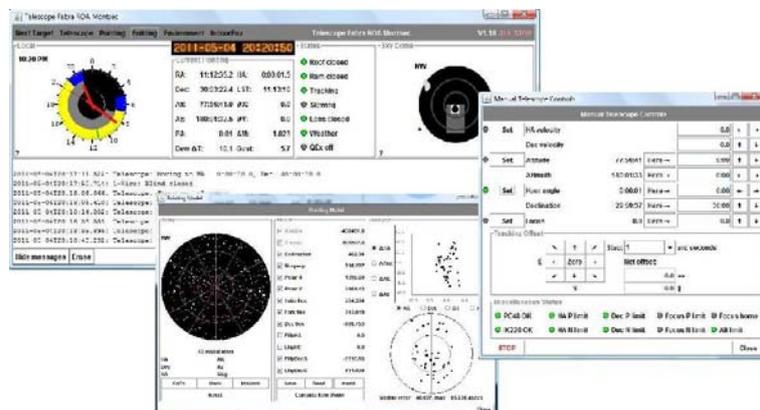

Fig.5. Some examples of the INDI-based control software windows in remote mode. Main window, control telescope and pointing model with up to 13 coefficients.

Three other Baker-Nunn cameras have been already mechanically refurbished, optically refigured and with CCD-based facilities: APT at Siding Springs [3], Phoenix at AMOS [4] and NESS-T at Rothney Astrophysical Observatory [5]. The experience of these previous efforts has significantly helped us to the successful completion of our refurbishment project. Later to our initiative, another BNC at ARIES (India) has been started to refurbish [6].

**FINAL INSTALATION AND COMMISSIONING**

In the summer of 2010, once the upgrade was finished in San Fernando and after a brief testing period of all the systems, the telescope was moved from ROA to its final observing site at the Montsec Astronomical Observatory (OAdM), whose WGS84 coordinates are; lat = 42.0516deg, long = 0.7293deg and height = 1622m HAE.
After installing the telescope with all its associated devices and with a good optical collimation achieved, the quality of the first technical images fully guaranteed the scientific usefulness of the telescope, as can be seen in Fig.6.

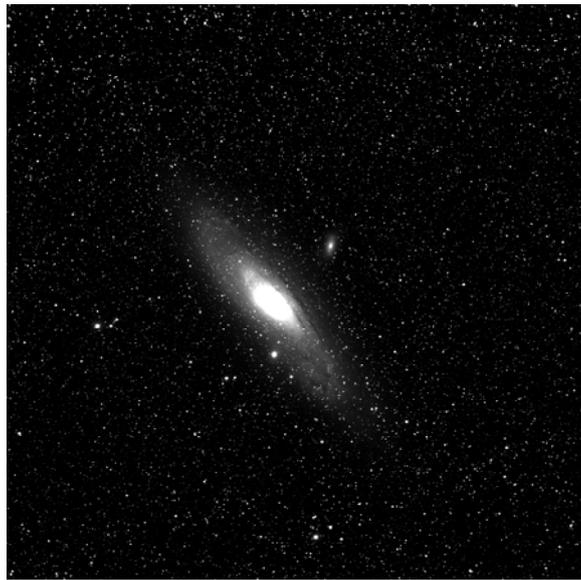
Fig.6. First 4.4º x 4.4º technical image of M31 taken at OAdM on September 11, 2010.

Finally, on September 16, 2010, the new facility was officially opened and renamed as Fabra-ROA Telescope at Montsec (TFRM). Fig.7 is a panoramic view of the OAdM with the TFRM in foreground. Note the blind that protects the objective lens, the sliding roof opened and the South wall end lowering.

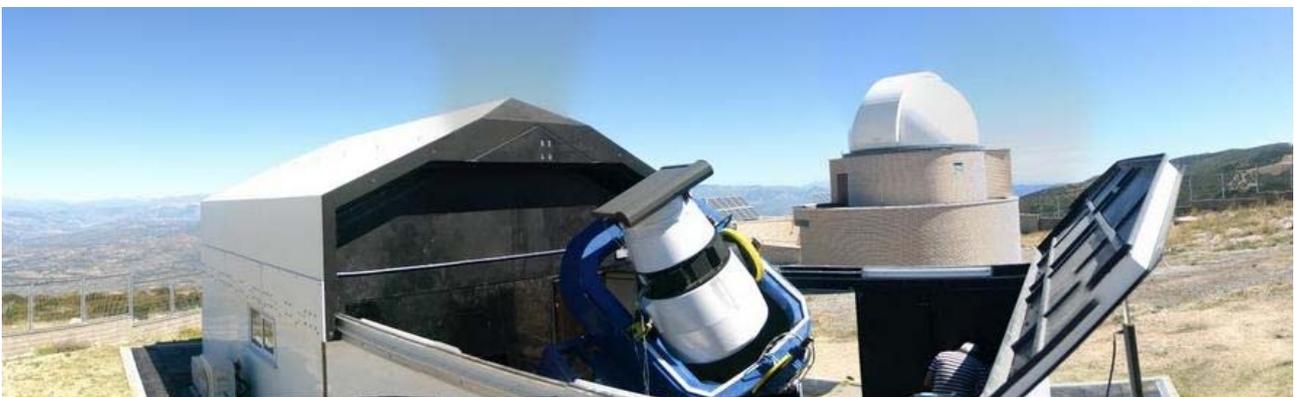
Fig.7. Astronomical Observatory of Montsec at Pre-Pyrenees Mountains, 1570m MSL height. Foreground, the TFRM inside its reinforced glass-fiber enclosure, with its sliding roof opened and the South wall end lowering.

**TFRM CAPABILITIES**

There are a significant number of features that makes TFRM a very suitable telescope in Space Surveillance and Tracking tasks, such as:

- To be a fully remote and robotic telescope. The INDI protocol [2] can handle all the telescope devices via internet (remote mode) or to schedule any kind of observation, e.g. interleave Space Debris and NEO observation task, without any operator intervention (robotic mode).
- The significant FoV, free of optical aberrations, that allows quick sky surveys.
- Its moderately deep limiting magnitude (V~20mag) with around 30s integration time.
- The HA and DEC motions at arbitrary rates that permit both, to track objects in any kind of orbits with just introducing the orbital elements and to carry out surveys of those orbits (LEO, MEO, GEO, etc.). Fig.8 shows an example of the INDI target window with a LEO satellite selected by its TLEs and Fig.9 shows some images details of three different satellites tracking in robotic mode.
- The camera shutter can be commanded at will during one single CCD integration, making possible to introduce time stamps in the image object trails.
- The timebase accuracy is assured to be well below the ms by a LANTIME M200 GPS time server.

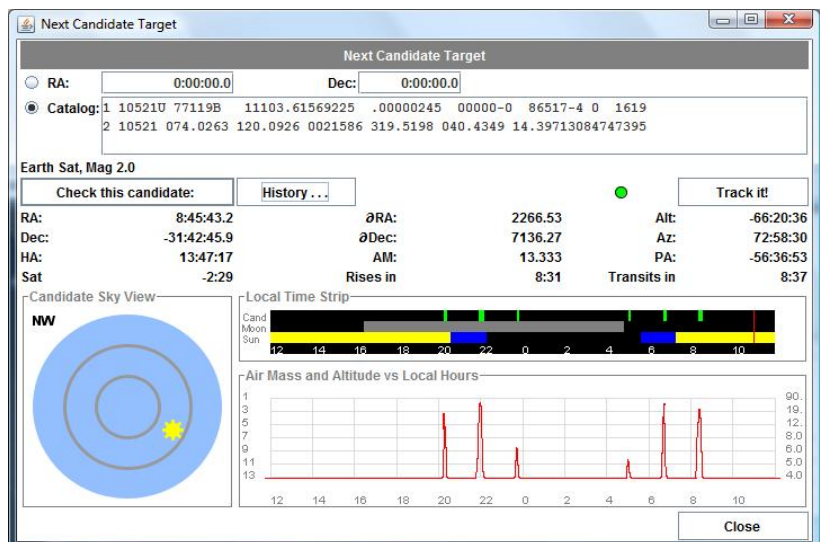

Fig.8. Target selection window. You can entry a celestial position by coordinates, catalogue name or by its orbital elements (TLEs or .edb format). The picture shows a tracking by TLE selection. Note the visibility Time Strip from the station and the logaritmical Air Mass vs. Local Time graph.

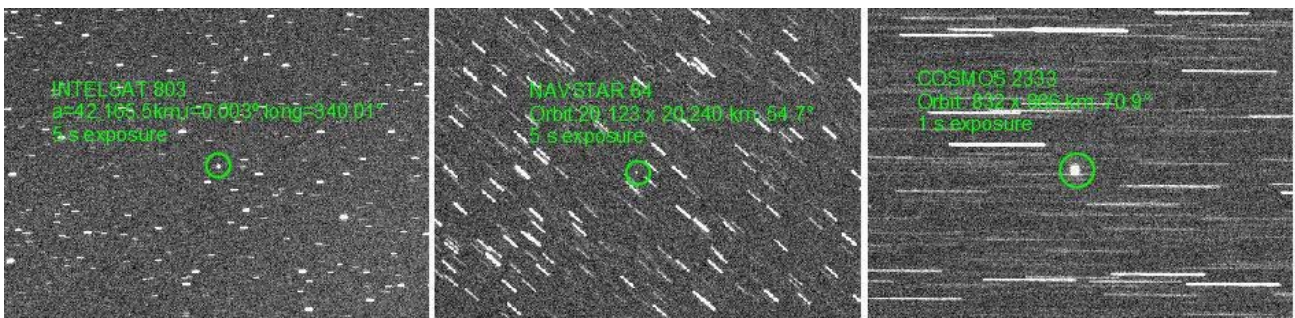

Fig.9. Three examples of satellite images taken with the TFRM while tracking in robotic mode. Note that the field of the images is only a small fraction of the 4.4º x 4.4º FoV. From left to right, a 5s exposure of the GEO satellite INTELSAT 803, a 5s exposure of one of the NAVSTAR/GPS constellation and a 1s exposure of the LEO satellite COSMOS 2333.

**FIRST TESTING RESULTS**

Although we were still in a commissioning period, the TFRM took part from January 31 to February 7, in the third optical observation campaign of GEO satellites, in the framework of the ESA Space Surveillance Awareness Preparatory Programme. In fact, this campaign has been the first observation task in which TFRM has participated. Systematic observations of different GEO satellites were conducted to determine 1137 satellite angular positions. The results will probably be presented by an oral communication in this congress. In any case, although our reduction software and the data pipeline are still being developed, we can estimate our astrometric precision in the GEO satellites angular coordinates to be below 0.5 arcseconds in both coordinates.

In order to testing our data quality, Orbit Determination (OD) from the angular measurements was carried out using the Orbit Determination Tool Kit (ODTK) software package, from Analytical Graphics, Inc. (AGI) [7]. This licensed software is used by ROA in collaboration with the Instituto Nacional de Técnica Aeroespacial (INTA). As an example, in Fig.10, we show the 2-sigma (95%) uncertainties obtained over the MSG2 satellite, with 175 angular measurements along 4 nights in which the satellite was not maneuvered.

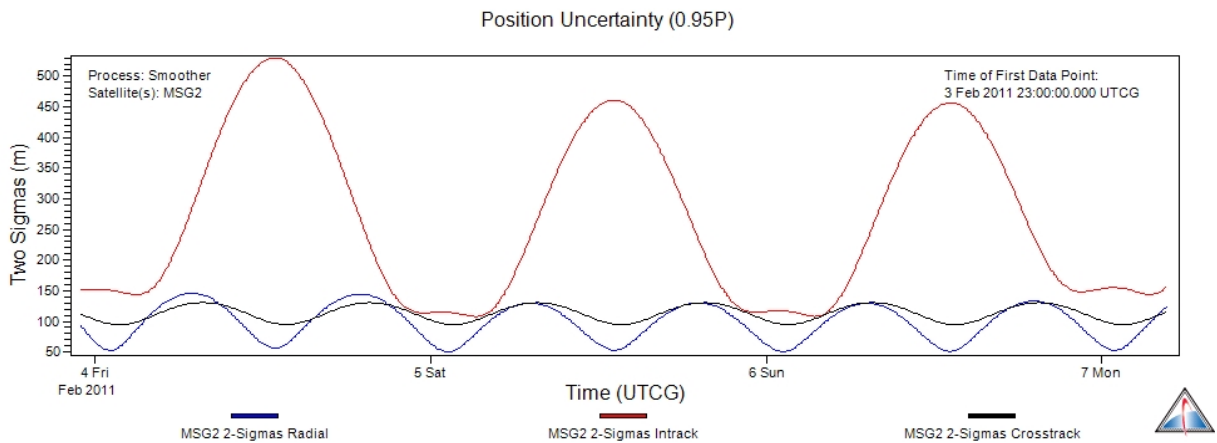

Fig.10. Uncertainties in the position of the MSG2 satellite, at 2 sigmas (95%) and in RIC coordinates. It is clearly noticeable the characteristic increasing of the Intrack uncertainty during daytime.

The mean uncertainties in the classical elements; semiaxis, eccentricity and inclination, are of the order of 12m, 1.8e-6 and 1.5e-4 deg respectively.

Although these results are quite preliminary reveal the usefulness of TFRM in GEO region tracking tasks.

On the last three nights of the ESA observation campaign (February 04 to 06 night), the NSS5 (INTELSAT 803) satellite was also observed from the ROA by the non-participant telescope Gautier Astrograph. In order to apply the reduction method developed by the PASAGE project [8], the ROA raw images has been reduced. The observations from ROA were very denses (one exposure every two minutes during all night, around 330 images per night). A simple inspection of the apparent topocentric orbits from the San Fernando station indicates a NSS5 maneuver on the February 04 to 05 night and a very clear E-W maneuver at the beginning of the last night, circles in Fig.11 Unfortunately we have only 75 TFRM observations of the NSS5 satellite in these three nights, therefore, the OD comparisons between ROA and TFRM data are not worthy.

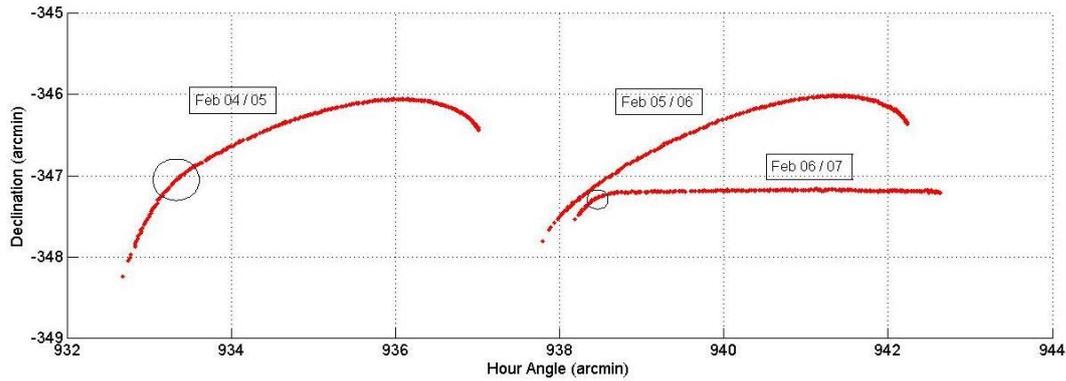

Fig.11. Apparent topocentric orbits of the NSS5 satellite from the San Fernando station, over the last three nights of the ESA observation campaign. From left to right, the February 04/05, 05/06 and 06/07 nights. Satellite positions are expressed in the Local TIRS Reference System, that is, a system fixed to Earth, instantaneous equator and the station meridian.

On February 18, 2011, the first unsupervised robotic mode observation for the TFRM was performed. It was the photometric monitoring of the gamma-ray source HESS J0632+057. Up to date, several known exoplanets, some gamma-ray sources counterparts and blazars, and dozens of asteroids have been observed in fully robotic mode (http://www.am.ub.es/bnc), in order to tune and test the system. We expect to continue with the TFRM commissioning period until the system will be in short fully reliable and automatic (including archiving and pipeline reduction).

**CONCLUSIONS**

Although the telescope system has multiple astronomical applications, we intend to devote a significant part of observing time to Space Surveillance and Tracking programmes. In fact, the TFRM is now one of the asset with which Spain contributes to the ESA for the future SSA/SST Programme.

Because of its particular features, TFRM is particularly useful in the surveillance and tracking of Space Debris and Near Earth Objects, thus we are looking forward to contribute to the international efforts in these fields.